# Broadband coherent Raman spectroscopy based on single-pulse spectral-domain ghost imaging

**Jing Hu,[1] Tianjian Lv,[1] Zhaoyang Wen,[1,2*] Wending Huang,[1] Ming Yan,[1,2*] Heping Zeng[1,2]**

[1]State Key Laboratory of Precision Spectroscopy, and Hainan Institute, East China Normal University, Shanghai 200062, China
[2]Chongqing Key Laboratory of Precision Optics, Chongqing Institute of East China Normal University, Chongqing 401120, China
*Corresponding author: zywen@lps.ecnu.edu.cn; myan@lps.ecnu.edu.cn;

**Broadband coherent anti-Stokes Raman scattering (CARS) spectroscopy plays a vital role in chemical sensing and label-free vibrational imaging, yet conventional methods suffer from limited acquisition speeds and complex detection schemes. Here, we demonstrate high-speed broadband CARS enabled by nonlinear spectral ghost imaging combined with time-stretch dispersive Fourier-transform spectroscopy (TS-DFT). We exploit modulation instability to generate a stochastic supercontinuum as the Stokes source and a synchronized narrowband pulse as the pump. Reference Stokes spectra are captured at 60.5 MHz via TS-DFT, while anti-Stokes signals are detected using a single non-spectrally resolving photodetector. Correlating these signals enables broadband CARS spectral reconstruction across the fingerprint (600–1600 cm$^{-1}$) and C–H stretching (2600–3400 cm$^{-1}$) regions with 13 cm$^{-1}$ resolution and microsecond-scale acquisition times. Our method enables robust signal recovery without the need for spectral resolution in the detection path, facilitating measurements in complex biological and chemical environments.**

CARS spectroscopy is a powerful nonlinear technique widely used for label-free chemical imaging and vibrational spectroscopy [1-3]. By leveraging the coherent interaction of laser fields with molecular vibrations, CARS provides high sensitivity, chemical specificity, and rapid acquisition speed, making it an indispensable tool in fields ranging from biomedical imaging [4] to material characterization [5]. However, emerging applications such as chemical kinetics [6], flow cytometry [7], and in vivo biological imaging [8] increasingly demand rapid acquisition of broadband spectra for multi-species detection—capabilities that remain challenging for conventional CARS implementations [9].

Ghost imaging, originally developed in the spatial domain, offers a fundamentally different approach to optical measurement by exploiting correlations between structured illumination patterns and a non-spatially resolving, bucket detector [10]. Recent advances have extended ghost imaging principles to the spectral domain, enabling spectroscopic detection without requiring direct spectral resolution at the detector [11,12]. This paradigm shift opens new possibilities for robust, high-dimensional spectral measurements with simplified detection architectures [13]. Despite these advances, conventional spectral ghost imaging (SGI) or stochastic covariance spectroscopy typically rely on programmable spectral filters [14] or spatial light modulators [15] to generate spectrally modulated illumination. The limited modulation rates of these devices impose fundamental constraints on measurement speed.

Here, we demonstrate high-speed, broadband CARS by integrating nonlinear SGI with TS-DFT. Our method harnesses modulation instability for generating stochastically modulated supercontinuum spectra. These reference spectra are captured at the pulse repetition rate (60.5 MHz) using TS-DFT spectroscopy. Concurrently, their nonlinear interaction with a sample generates CARS signals, which are recorded using a non-spectrally resolving single-point detector. By correlating TS-DFT spectra with CARS signal intensities, we reconstruct broadband coherent Raman spectra, spanning 600-1600 cm$^{-1}$ and 2600-3400 cm$^{-1}$ (resolution of 13 cm$^{-1}$), with microsecond-scale acquisition times.

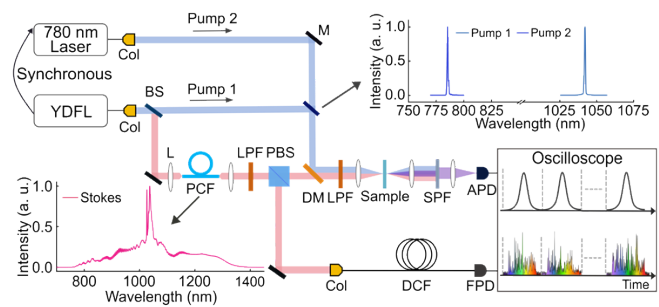

**Fig. 1.** Experimental setup. YDFL, ytterbium-doped fiber laser; Col, collimator; BS, 50:50 beam splitter; M, mirror; L, lens; PCF, photonic crystal fiber; LPF, long-pass optical filter; PBS, polarizing beam splitter; DM, dichroic mirror; SPF, short-pass optical filter APD, avalanche photodiode; DCF, dispersion-compensating fiber; FPD, fast photodetector.

Our experimental setup is illustrated in Fig. 1. A custom-built ytterbium-doped fiber laser (YDFL) operating at a 60.5 MHz repetition rate is split into pump and Stokes beams. The pump pulses are centered at 1042 nm with a spectral bandwidth of 0.84 nm (8 cm$^{-1}$), a duration of ~3 ps, and an average power of 100 mW. The Stokes pulses are compressed to 190 fs using a transmission grating pair and then coupled into a 70-cm-long photonic crystal fiber (PCF) via an aspheric lens, generating a supercontinuum spanning 800 to 1400 nm. After filtering with a long-pass filter (1100 nm cutoff), the Stokes spectrum exhibits a 3-dB bandwidth of 200 nm, covering Raman shifts from 600 to 1600 cm$^{-1}$. Finally, the Stokes pulses are further compressed to 90 fs using a prism pair, with an average power of 70 mW.

For SGI measurements, the Stokes pulses are split into reference (10%) and test (90%) arms via a beam splitter. In the reference arm, the Stokes pulses are temporally stretched by propagating through a 500-m dispersion compensating fiber (DCF) with a total group velocity dispersion of 28.4 ps·nm$^{-1}$. The stretched pulses are detected by a fast photodetector (20 GHz bandwidth), and the output is digitized using an 8-bit oscilloscope (33 GHz bandwidth, 40 GSa/s sampling rate). In the test arm, the pump (angular frequency: $\omega_p$) and Stokes ($\omega_s$) beams are combined via a dichroic mirror and filtered by a long-pass filter (1000 nm cutoff) to eliminate the short-wavelength background. The combined beams are then focused onto a 1-mm sample cuvette using an aspheric lens (f=19 mm). The forward-propagating anti-Stokes signal ($\omega_a$), generated via resonant four-wave-mixing process ($\omega_a=2\omega_p-\omega_s$), is collected by a second lens and spectrally purified by a short-pass filter (SPF, 1000 nm cutoff) to block residual pump and Stokes light. Finally, the signal is detected by an avalanche photodiode (APD, 1 GHz bandwidth) and digitized with the same oscilloscope.

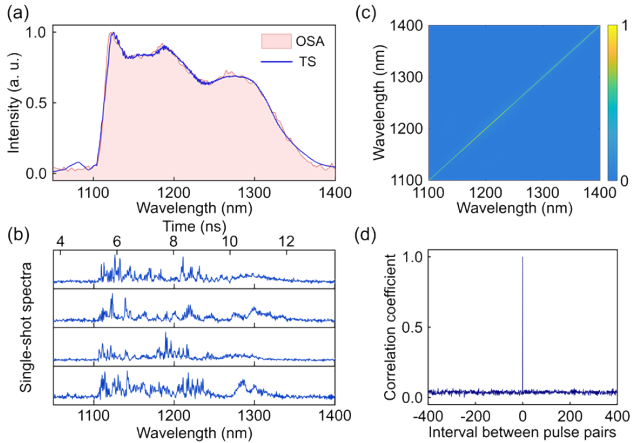

**Fig. 2.** Characterization of Stokes supercontinuum. (a) Comparison of time-stretch spectra with a steady-state spectrum obtained using an optical spectrum analyzer (OSA). (b) Four single-shot spectra acquired via the DFT technique. (c) Wavelength-to-wavelength Pearson correlation matrix for the filtered broadband Stokes spectrum, calculated from an ensemble of 5000 single-shot measurements. (d) Cross-correlation between the test and reference pulse pairs as a function of their temporal interval N.

In linear SGI, the spectral information can be reconstructed via the normalized correlation function $C(\lambda)$, defined as [11]:

$$C(\lambda) = \frac{<\Delta I_{ref}(\lambda) \cdot \Delta I_{test}>_N}{\sqrt{<[\Delta I_{ref}(\lambda)]^2>_N <[\Delta I_{test}]^2>_N}},$$

where $<>_N$ denotes ensemble averaging over N measurements, $\Delta I = I - I_{mean}$ represents intensity fluctuations relative to the mean, $\Delta I_{ref}(\lambda)$ and $\Delta I_{test}$ are fluctuations of the reference spectra and the measured CARS intensities, respectively. Since CARS is a third-order nonlinear process, the anti-Stokes signal ($I_a \propto I_s \cdot I_p^2$) depends linearly on the Stokes intensity ($I_s$) and quadratically on the pump intensity ($I_p^2$). Therefore, in our measurements, the pump intensity is held constant, while only the Stokes beam is spectrally modulated. As such, the system adheres to the linear correlation framework described above.

Fig. 2 presents the recorded reference spectra. For TS-DFT measurements, we first perform time-to-wavelength calibration [16,17]. As shown in Fig. 2(a), the calibration is established by comparing the TS-DFT results (6000-fold averaging, blue curve) with a reference spectrum (red curve) measured with an optical spectrum analyzer (OSA; Yokogawa, AQ6375; resolution: 0.1 nm). The excellent agreement between these spectra validates our TS-DFT spectrometer. We then measure stochastic single-shot Stokes spectra at the pulse repetition rate of 60.5 MHz (corresponding to 16.5 ns per spectrum). The results exemplified in Fig. 2(b) exhibit significant shot-to-shot variations due to modulation instability in supercontinuum generation [18,19]. To further characterize the stochastic nature of the modulated spectra, we calculate the wavelength-to-wavelength Pearson correlation matrix using 5,000 single-shot spectra, as shown in Fig. 2(c). The matrix exhibits near-zero correlation coefficients across all wavelength pairs, except along the diagonal (where $\lambda_1 = \lambda_2$). This absence of off-diagonal correlations confirms the spectral randomness, which is essential for spectral-domain ghost imaging [20]. Furthermore, to maintain temporal correlation between the reference and test beams, we add an electronic delay line in the test arm (positioned between the APD and oscilloscope) to compensate for the propagation delay induced by the 500-m DCF in the reference arm. Fig. 2(d) presents the cross-correlation measurement of 10000 pulse pairs as a function of pulse number interval (N), showing a correlation peak occurring only at N=0. This zero-delay peak confirms that correlated pulses exclusively originate from the same beam-splitting event.

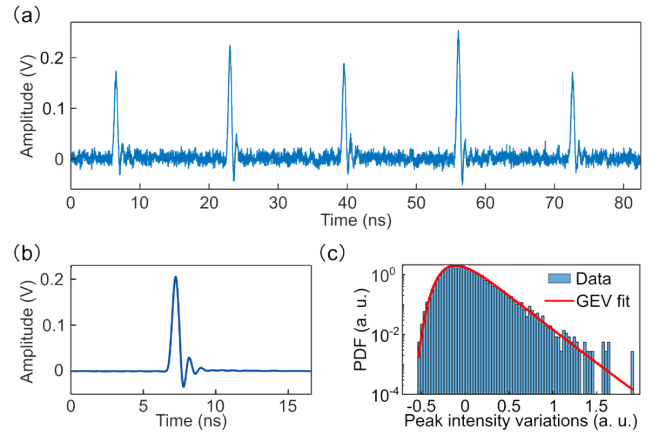

**Fig. 3.** Analysis of anti-Stokes pulse intensity variations. (a) Anti-Stokes pulses acquired in real time. (b) Averaged anti-Stokes signal within a 200 μs acquisition window. (c) Probability density function (PDF) of peak intensity fluctuations relative to the mean, calculated from 12,000 individual anti-Stokes pulses, fitted with a generalized extreme value (GEV) distribution (red solid line).

In a proof-of-principle demonstration, we measure Raman fingerprints of pure toluene. The anti-Stokes pulses recorded in the test arm (Fig. 3(a)) display intensity fluctuations that correlate with the corresponding reference spectra. These pulses exhibit a mean signal-to-noise ratio (SNR) of approximately 30, calculated as the peak intensity divided by the standard deviation (SD) of the noise floor. By applying 12000 averages (200-μs total acquisition time), the SNR improves significantly to 400, as shown in Fig. 3(b). To further characterize the signal statistics, we evaluate the probability density function (PDF) of 12000 anti-Stokes pulses, normalized to their mean value. The histogram of these fluctuations (Fig. 3(c)) closely follows a generalized extreme value (GEV) distribution (red solid line) [21].

Next, we use the normalized correlation function between the CARS signals and the reference TS-DFT spectra to reconstruct the CARS spectrum. We calculate the Raman shifts using the pump and Stokes' wavenumbers. The reconstructed spectra, obtained within 5 μs, 100 μs and 200 μs, respectively, are shown in Fig. 4(a). The Raman signatures corresponding to the vibrations of the benzene ring at 786 cm$^{-1}$, 1003 cm$^{-1}$, and 1210 cm$^{-1}$, respectively, are observed. For these spectra, the non-resonant background is removed using the widely-used maximum entropy method (see the details in our previous works [22]). In Fig. 4(a), we measure a SNR of 12 for the 1003 cm$^{-1}$ Raman line with a 5 μs acquisition time. By extending the integration time to 200 μs (12000 averaging events), the SNR improves to 77, consistent with the expected √N scaling for noise reduction in spectral reconstruction, as indicated in Fig. 4(b). We determine the spectral resolution to be 13 cm$^{-1}$, obtained from the full width at half maximum (FWHM) of the dominant 1003 cm$^{-1}$ Raman peak. This experimental value agrees well with the theoretically predicted 12 cm$^{-1}$ resolution derived from TS-DFT analysis, which is governed by the 20 GHz bandwidth of the fast photodetector. Additionally, a measured spectrum for pure acetone is exemplified in Fig. 4 (c), showing the Raman lines at 790 cm$^{-1}$ and 1070 cm$^{-1}$.

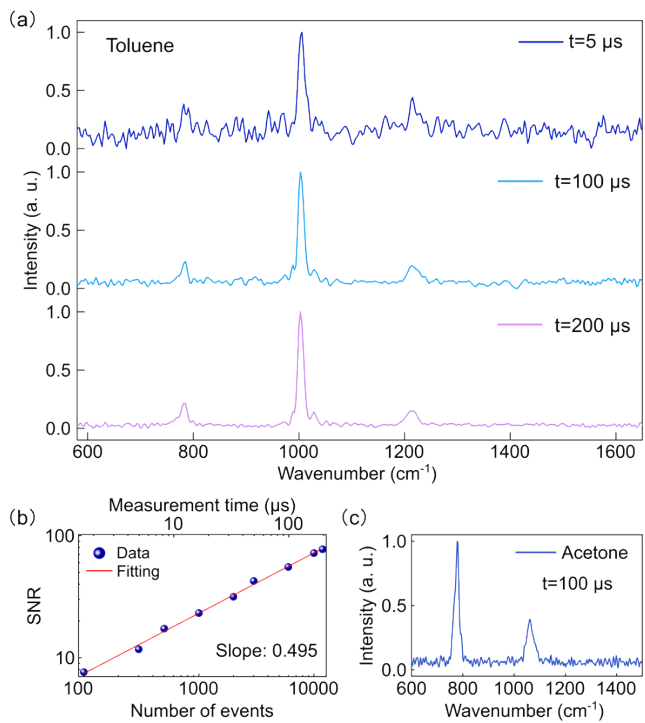

**Fig. 4.** Results of CARS spectra measurements in the fingerprint region. (a) CARS spectrum of toluene acquired with different averaging times. (b) SNR of the 1000 cm$^{-1}$ Raman peak as a function of the measurement time. (c) Broadband CARS spectrum of pure acetone.

Finally, we demonstrate our method in the high-wavenumber region (2600-3400 cm$^{-1}$) for characterizing C-H stretching modes. To this end, the 1042-nm pump beam is replaced with the output of a synchronized mode-locked laser (center wavelength: 785 nm; bandwidth: 0.8 nm; average power: 50 mW; pulse duration: 1 ps). Meanwhile, the short-wavelength region (950-1100 nm) of the supercontinuum serves as the Stokes beam. Correspondingly, the 1100-nm long-pass filter is substituted with an 1100-nm short-pass filter. Also, the 1000-nm long-pass filter is replaced by a 780-nm long-pass filter to block residual pump light, and the 1000-nm short-pass filter is changed to a 750-nm short-pass filter to isolate the anti-Stokes signal. All other experimental components remain unchanged.

Fig. 5(a) compares the time-averaged TS-DFT spectrum of the Stokes beam with a steady-state spectrum obtained from OSA, demonstrating excellent agreement in the short-wavelength region and validating the accuracy of our TS-DFT system. In Fig. 5(b), we present the Pearson correlation matrix derived from 5,000 independent spectral measurements. The matrix shows strong correlations exclusively along the diagonal, while off-diagonal coefficients are statistically negligible (approaching zero), which confirms effective spectral decorrelation and minimal cross-talk between different spectral components. In this experiment, we measure the high-wavenumber C-H stretching vibrations of toluene. Fig. 5(c) displays the reconstructed coherent Raman spectrum covering the 2800-3400 cm$^{-1}$ region, revealing two prominent Raman bands at 2915 cm$^{-1}$ and 3052 cm$^{-1}$. The 3052 cm$^{-1}$ Raman mode demonstrates strong signal performance, achieving a SNR of 40 with a 100 μs acquisition time (note that the SNR follows the expected √t scaling with acquisition time).

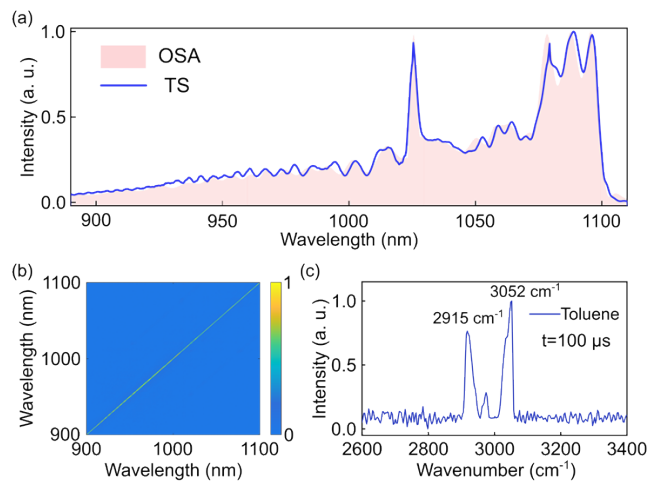

**Fig. 5.** Spectral measurements in the high-wavenumber region. (a) Time-stretch DFT spectra compared with optical spectral analyzer (OSA) measurements. (b) Wavelength-to-wavelength correlations of the filtered broadband spectrum measured across 5,000 single-shot spectra. (c) CARS spectrum of toluene reconstructed via spectral ghost imaging, spanning 2600-3400 cm$^{-1}$.

Notably, recent advances in broadband CARS have dramatically improved measurement speeds. For example, dual-comb

spectroscopy—enabled by two asynchronous optical frequency combs and a fast photodetector—has achieved high-speed coherent Raman measurements across both the fingerprint (200–1400 cm$^{-1}$) [23–26] and C–H stretching (2800–3600 cm$^{-1}$) [27,28] regions, with sub-10 cm$^{-1}$ resolution and microsecond acquisition times. Similarly, nonlinear TS-DFT spectroscopy has demonstrated broadband coherent Raman spectral measurements (200–1200 cm$^{-1}$) at the laser repetition rate (50 MHz), setting new speed records for vibrational spectroscopy [29]. However, these methods depend on either complex laser systems or intricate detection schemes, restricting their practical adoption.

In contrast, SGI provides a compelling alternative for CARS by using a simplified detection scheme with a non-spectrally-resolving detector. Like its spatial and temporal counterparts, SGI offers inherent advantages such as high photon efficiency and robustness to scattering—critical for low-light applications like deep-tissue imaging. For instance, our system detects anti-Stokes signals (Fig. 3a) with just nanowatt-level average power, limited only by the current detector's sensitivity (further improvable with a photomultiplier tube). While SGI leveraging supercontinuum modulation instability and TS-DFT has been applied to linear gas spectroscopy [11], its extension to nonlinear spectral domains had remained unexplored. Our work now bridges this gap, unveiling the potential of TS-DFT-based SGI for high-speed nonlinear molecular fingerprinting.

In conclusion, we demonstrate high-speed nonlinear SGI for broadband CARS measurements. By leveraging TS-DFT spectroscopy, we achieve Stokes spectrum acquisition at refresh rates up to 60.5 MHz, enabling rapid reconstruction of Raman fingerprints across both the fingerprint (600-1600 cm$^{-1}$) and C-H stretching (2600-3400 cm$^{-1}$) regions with 13 cm$^{-1}$ spectral resolution and microsecond-scale integration times. Since spectral detection is eliminated in the test arm, our approach offers distinct advantages for low-photon-flux measurements and turbid samples like biological tissues. This capability opens new possibilities for real-time vibrational imaging in biological systems and advanced material characterization.

**Funding.** This work was financially supported by the Innovation Program for Quantum Science and Technology (2023ZD0301000) and the Research Project of Shanghai Science and Technology Commission (22560730400).

**Disclosures.** The authors declare no conflicts of interest.

**Data availability**. Data underlying the results presented in this paper are not publicly available at this time but may be obtained from the authors upon reasonable request.